\documentclass[checkin,showpacs,aps,pra]{revtex4-1}

\newcommand{\neel}{N$\acute{\rm e}$el }

\newcommand{\bn}{\begin{enumerate}}
\newcommand{\en}{\end{enumerate}}
\newcommand{\ba}{\begin{eqnarray}}
\newcommand{\ea}{\end{eqnarray}}

\usepackage{graphicx,color}

\newcommand{\be}{\begin{equation}}
\newcommand{\ee}{\end{equation}}

\newcommand{\et}{{\it et al. }}

\def\prl{{ Phys. Rev. Lett. }}

\begin{document}

\newcommand{\clr}{}








\title{All-optical spin switching under different spin configurations}


\author{G. P. Zhang$^{*}$ and Mitsuko Murakami}

 \affiliation{Department of Physics, Indiana State University,
   Terre Haute, IN 47809, USA }

\date{\today}

\begin{abstract}
{ All-optical spin switching represents a new frontier in
  femtomagnetism. However, its underlying principles are quite
  different from traditional thermal activated spin switching.  {Here,
    we employ an atomic spin model and present a systematic
    investigation from a single spin to a large system of over a
    million spins.}  We find that {for a single spin without an
    external perturbation, the conservation of total angular momentum
    requires that the spin change, if any, exactly matches the orbital
    momentum change,} but a laser pulse significantly alters this
  relation, where the spin change does not necessarily follow the
  orbital change. This is reflected in the strong dependence of
  switching on laser polarization. To have an efficient spin
  switching, the electron initial momentum direction must closely
  follow the spin's orientation, so the orbital angular momentum is
  transverse to the spin and consequently the spin-orbit torque lies
  in the same direction as the spin. The module of the spin-orbit
  torque is $\lambda |{\bf S}||{\bf r}||{\bf P}|
  \sqrt{\cos^2\alpha+\cos^2\beta-2\cos\alpha \cos\beta \cos\gamma} $,
  where $\alpha(\beta)$ is the angle between spin {\bf S} and position
  {\bf r}(momentum { \bf P}) and $\gamma$ is the angle between {\bf r}
  and {\bf P}. These findings are manifested in a much larger system.
  We find that the spin response depends on underlying spin
  structures. {A linearly polarized laser pulse creates a dip in
    a uniform inplane-magnetized thin film, but has little effects on
    \neel and Bloch walls. Both right- and left- circularly polarized
    light ($\sigma^+$ and $\sigma^-$) have stronger but different
    effects in both uniform spin domains and \neel walls.}  While
  $\sigma^+$ light creates a basin of spins pointing down, $\sigma^-$
  light creates a mound of spins pointing up. In the vicinity of the
  structure spins are reversed, similar to the experimental
  observation.  $\sigma^+$ light has a dramatic effect, disrupting
  spins in Bloch walls. {By contrast, $\sigma^-$ light has a
    small effect on Bloch walls because $\sigma^-$ only switches down
    spins up and once the spins already point up, there is no major
    effect}.  These findings are expected to have important
  implications in the future.  }
\end{abstract}




 \maketitle


\section{Introduction}

Manipulating electron spins in magnetic media with light is
interesting, because it underlies the technology behind
magneto-optical recording in rare-earth and transition-metal alloys
\cite{kryder1985}. After a laser beam heats a sample over its
compensation point, a magnetic field is applied to switch the spin to
a different direction \cite{ourreview,rasingreview}.  It is also known
that if the sample has a compensation temperature slightly higher than
ambient temperature, the demagnetizing field plays a role as a bias
field for thermomagnetic writing, free of a magnetic field
\cite{shieh1986}. This is an earlier version of all-optical spin
switching (AOS), but it did not touch the time scale involved. In
2007, the same material showed AOS but on a much faster time scale
\cite{stanciu2007}.  To this end, a large number of materials have
been found to show AOS \cite{mangin2014}, besides rare-earth
transition metal alloys \cite{alebrand2012}. Many synthesized alloys
show the same behavior, and recently even a Pt/Co/Pt ferromagnetic
stack showed AOS \cite{vomir2017}. Parlak \et \cite{parlak2018}
further demonstrated AOS in CoPt multilayers, which can be further
optimized \cite{yamada2019}.

However, the microscopic mechanism of AOS is still under debate (see
reviews \cite{mplb16,mplb18}).  The proposed mechanisms generally fall
into two categories: Helicity-dependent (HD) and helicity-independent
(HID) switching. {Helicity refers to how the electric field of
  light rotates with respect to the light propagation
  direction. Figure \ref{fig1} shows an example of left-circularly
  polarized light ($\sigma^-$).}  It was originally thought that the
compensation temperature is critical to ultrafast AOS just as that in
slow AOS, but this is not always the case.  Hassdenteufel \et
\cite{hassdenteufel2013} and Schubert \et \cite{schubert2014a} showed
that switching can appear below or above the magnetic compensation
point or in a sample without a compensation point.  These two types of
switching can appear in the same material, depending on the laser
field fluence \cite{ostler2012}.  Experimentally when a laser pulse
scans across their magnetic sample, at the center of laser excitation
the domain is demagnetized, with randomly distributed spin up and down
domains, and the spin switching appears in the vicinity of the
demagnetized region \cite{stanciu2007}. We recently showed that when
one has a ferrimagnet \cite{prb17}, care must be taken because two
different helicities act upon two respective sets of spins, leading to
apparent HID switching.  Because optical selection rules depend on the
spatial orientation of spins in a domain \cite{jpcm17a,jpcm17c,prb17},
even subject to the same laser beam, spins inside the domain wall
behave differently. This was observed experimentally in multiferroic
TbMnO$_3$ \cite{manz2016} which shows a significant dependence on spin
configuration.  Manz \et \cite{manz2016} demonstrated a chirality
reversal of multiferroics by a linearly polarized laser pulse which
induces a skymion structure through magnetoelectric interactions.

In this paper, we aim to carry out a systematic investigation as how
system parameters and laser parameters affect all-optical spin
switching.  We employ a model that takes into account the spin-orbit
coupling, spin-spin interaction and the interaction between the laser
field and system, without resorting to an effective magnetic field
\cite{ostler2012}. We start with a single spin to demonstrate that a
laser pulse can significantly decouple the spin change from the
orbital angular momentum change. We find that rather than through
momentum exchange with the orbital, the spin change is better
described by the spin-orbit torque. This is revealed by using
different laser polarizations and initial spin orientations. By
changing the direction of momentum, we find that to reverse spins, it
is necessary to have a momentum along the spin direction, so the
transverse momentum is boosted to generate a large longitudinal torque
on the spin. These intricate relations are manifested in a thin film
of a million spins. We employ three different types of magnetic
domains: a uniform inplane magnetized domain, \neel and Bloch walls.
The same linearly polarized light induces very different changes. The
uniform magnetized film shows a dip at the center of excitation, but
\neel and Bloch walls show a much weaker change.  Circularly polarized
light has a much stronger effect. We find that circularly polarized
light can directly write information into a uniform spin domain, \neel
and Bloch walls. A circular pattern is formed and around the vicinity
of the pattern, the spins are reversed, similar to the experimental
results \cite{stanciu2007}. At the center of excitation, spins are
disoriented and chaotic, also similar to the experiment
\cite{stanciu2007}.  Right- ($\sigma^+$) and left- ($\sigma^-$)
circularly polarized light have different effects. $\sigma^+$ tends to
flip spins from up to down, while $\sigma^-$ does the opposite. This
explains why $\sigma^-$ does not change the Bloch wall strongly.  Our
findings shed fresh light on AOS and are expected to have important
impacts on future experimental and theoretical investigations.

{The rest of the paper is arranged as follows. In Sec. II, we
  outline our theoretical formalism. Section III is devoted to the
  results and discussions. We present our findings on a single spin
  first and then on domains of three different types, uniform domains
  with inplane magnetization in Sec. III(C), \neel walls in
  Sec. III(D) and Bloch walls in Sec. III(E). We provide an analytic
  theory on spin-orbit torque in Sec. III(F).}  Finally, we conclude
this paper in Sec. IV. An appendix is provided to discuss the role of
orbital angular momentum.

\section{Theoretical formalism}

\begin{figure}
  \includegraphics[angle=0,width=1\columnwidth]{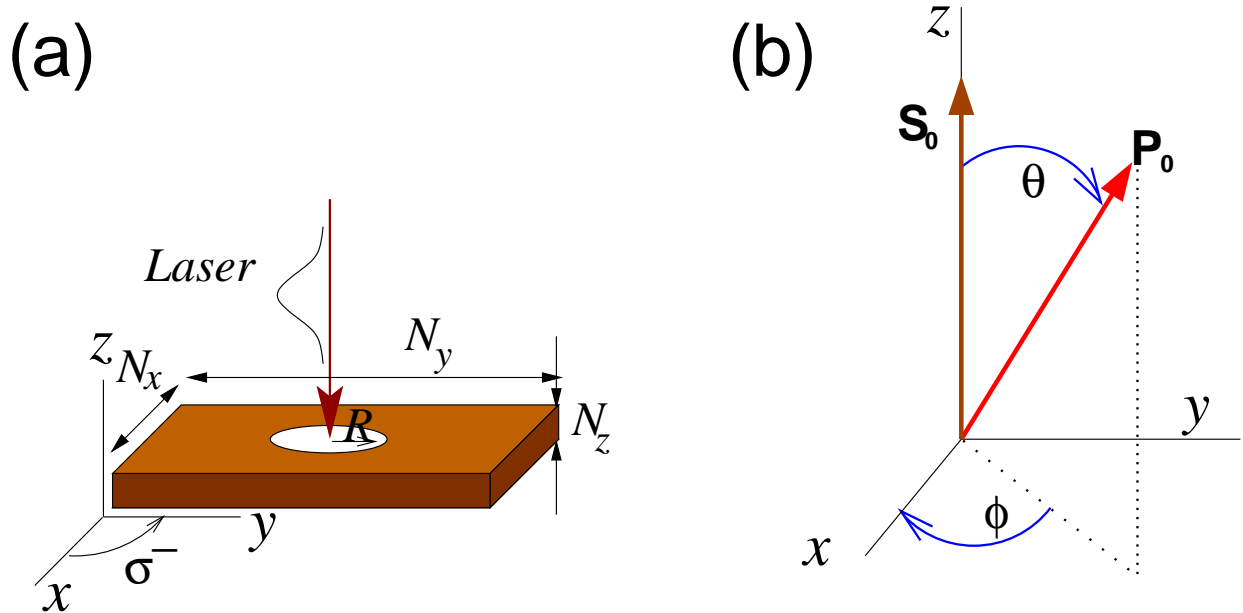}
  \caption{(a) Schematic of our simulated magnetic thin film, with
    dimension $N_x\times N_y\times N_z$, where $N_x=N_y=501$ and
    $N_z=4$.  A laser pulse impinges on the sample with radius $R$,
    and its polarization can be linear or circular. {In the
      bottom of the figure, we show one example of left-circularly
      polarized light ($\sigma^-$). In our convention, we use the
      right hand rule with the thumb along the light propagation
      direction, and if the electric field follows the direction of
      the fingers, the light is right circularly polarized, otherwise,
      it is left-circularly polarized light. } (b) Definition of the
    spatial orientation $(\theta,\phi)$ of the initial momentum ${\bf
      P}_0$ with respect to the initial spin ${\bf S}_0$ for the
    single spin calculation. In many-spin calculations, ${\bf S}_0$
    can lie inplane or have any particular orientation.  }
\label{fig1}
  \end{figure}

All-optical spin switching employs an ultrafast laser pulse to switch
spins in a magnetic material. Figure \ref{fig1}(a) represents a
typical experimental geometry with lattice sites $(N_x\times N_y
\times N_z)$, where the pulse is propagating along the $-z$ axis and
illuminates a small area of the sample with radius $R$.  The pulse can
be either linear or circularly polarized. The magnetic films contain
many spins, with various domain structures.  Any reasonable theory
must treat both the laser field and system properly.  AOS is an
optical process.  Under dipole approximation, light does not interact
with spin because the selection rule dictates spin conservation, or
$\Delta S=0$. This naturally creates a major problem for theoretical
treatment. To circumvent this difficulty, one often introduces a
spin-symmetry breaking term such as an effective magnetic field
\cite{ostler2012} to simulate laser pulses. This effective field
should not be confused with the laser's own magnetic
field. First-principles calculations are possible, but are unable of
reproducing spin reversal \cite{chimata2015,prb17}.

When we were investigating the magneto-optical effect, we accidentally
discovered a model \cite{epl15,epl16} that allows us to simulate spin
reversal, without an effective magnetic field. Different from other
studies, we augment a spin-orbit coupling term, $\lambda {\bf
  L}_i\cdot {\bf S}_i$, to the Heisenberg exchange model, where
$\lambda$ is the spin-orbit coupling constant, ${\bf L}_i$ and ${\bf
  S}_i$ are the orbital and spin angular momenta at site $i$.  Our
Hamiltonian is \cite{epl15,epl16,jpcm17a,jpcm17b} \be H=\sum_i \left ( \frac{{\bf P}_i^2}{2m}+V({\bf r}_i)
+ \lambda {\bf L}_i\cdot {\bf S}_i -e {\bf E}(t) \cdot {\bf
  r}_i \right ) -J\sum_{ij} {\bf S}_i\cdot {\bf S}_j
\label{ham} \ee where the first term is the kinetic energy
operator of the electron, the second term is the potential energy
operator, and the third term is the spin-orbit coupling. The fourth
term in Eq. \ref{ham} describes the interaction between the laser
pulse ${\bf E}(t)$ and the system \cite{epl15}, so we do not have to
employ an effective magnetic field approximation. {The laser
  field, centered with respect to magnetic domains, propagates
  vertically down along the $-z$ axis, with the electric field \be
  {\bf E}({\bf r},t)={\bf
    A}(t)\exp[-\frac{(x-x_c)^2+(y-y_c)^2}{R^2}-\frac{z}{d}], \ee where
  $x$ and $y$ are the coordinates in the unit of the site number,
  $x_c$ and $y_c$ denote the center location of the laser spot, $R$ is
  the radius of the laser spot ($R$ is 50 lattice sites in our
  simulation, see Fig. \ref{fig1}(a)), and $d$ is the penetration
  depth of light ($d=30$ lattice sites in our case).  For a left
  (right)-circularly polarized field [$\sigma^-$($\sigma^+$)], ${\bf
    A}(t)$ is \be {\bf A}(t)=A_0{\rm e}^{-t^2/T^2}\left
  [\mp\sin(\omega t) \hat{x}+\cos(\omega t) \hat{y} \right ], \ee
  where $\omega$ is the laser carrier frequency, $T$ is the laser
  pulse duration, $A_0$ is the laser field amplitude, $t$ is time, and
  $\hat{x}$ and $\hat{y}$ are unit vectors, respectively.  For a
  linearly polarized field ($\pi$), ${\bf A}(t)$ is \be {\bf
    A}(t)=A_0{\rm e}^{-t^2/T^2}\cos(\omega t) \hat{x}.  \ee These
  three fields are used in the following calculation.  The laser pulse
  duration $T$ is 60 fs and the field amplitude takes several
  different values (see below for details).  Figure \ref{fig1}(a)
  shows our simulation box with $N_x=501$ spins along the $x$ axis,
  $N_y=501$ spins along the $y$ axis, and $N_z=4$ spins along the $z$
  axis.  We take into account three types of spin configuration: (i)
  uniform spin domains, (ii) Bloch and (iii) \neel walls.  The domain
  wall width for both \neel and Bloch walls is 200 lattice sites.}
The last term in Eq. (\ref{ham}) is the exchange coupling between two
neighboring spins, and $J$ is the exchange parameter.  {We choose
  $S=1\hbar$, $J=0.1$ eV$/\hbar^2$, $\lambda=0.06$ eV$/\hbar^2$
  \cite{prl00}}.  We adopt the harmonic oscillator potential for
$V({\bf r})$.  This Hamiltonian is similar to the traditional harmonic
oscillator model for electron in magneto-optics \cite{epl15}.

 To compute spin dynamics, we solve the equation of motion for each
 operator $O$ of interest, $i\hbar \dot{O}=[O,H]$.  For instance, the
 spin changes according to \be \frac{d{\bf S}_i}{dt}= \lambda {\bf
   L}_i\times {\bf S}_i +J \sum_j{\bf S}_i\times {\bf S}_j.
\label{spin0}
\ee {Since our main interest is to compute a system with many
  spins, we make a Hartree-Fock approximation to the coupled operator,
  where operators are replaced by their expectation values.  More
  accurate calculations are possible, but then one is restricted to a
  system size too small to compare with experimental sizes. This is
  the compromise that we have to make.}
\section{Results and discussions}

To develop a physical picture for AOS, we start with a single spin and
then move on to more complex systems with a million exchange-coupled
spins. The results on a single spin serve the basis for our research.

\subsection{Single spin}

\begin{figure}
  \includegraphics[angle=270,width=1\columnwidth]{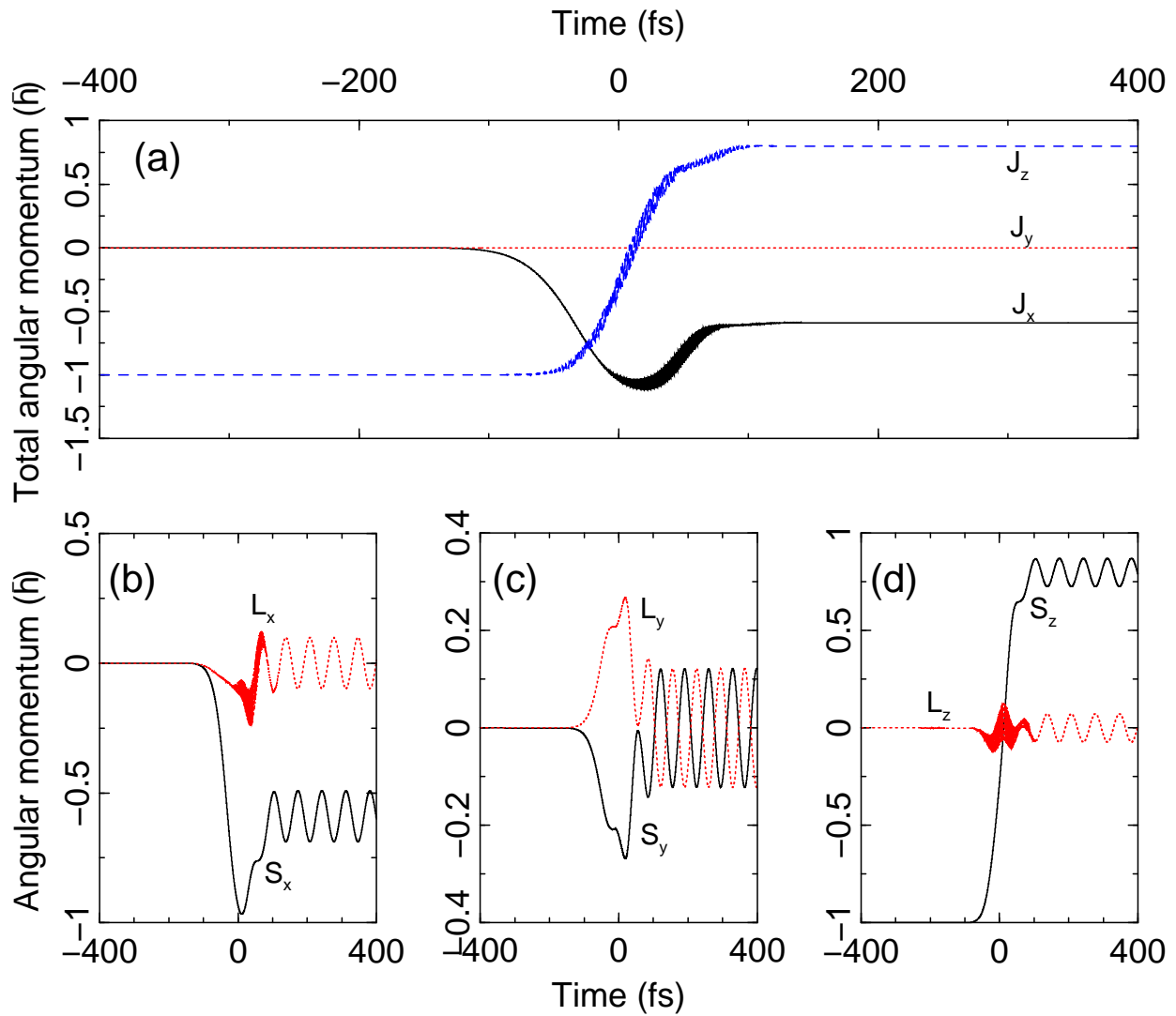}
  \caption{ (a) Total angular momentum {\bf J} in a single spin as a
    function of time. Upon laser excitation, it changes rapidly.
    Before and after laser excitation, it is conserved. (b) The spin
    and orbital angular momentum along the $x$ axis. They are
    decoupled once the system is excited.  (c) and (d) Same as (b) but
    for the $y$ and $z$ axes, respectively.  Note that $L_z$ and $S_z$
    are very different.  }
\label{singlespin}
\label{fig2}
  \end{figure}

For a single spin, no exchange interaction is present and only the
spin-orbit coupling remains.  Under influence of spin-orbit coupling,
the spin at site $i$ precesses according to \be \frac{d{\bf S}_i}{dt}=
\lambda {\bf L}_i\times {\bf S}_i.
\label{spin}
\ee
The right hand side is the spin-orbit torque, and depends on the
orbital and spin itself. 
 The orbital angular momentum precesses as
\be
\frac{d{\bf L}_i}{dt}=- \lambda {\bf L}_i\times {\bf S}_i.
\label{orbital}
\ee So one can see that for a single spin system, ${\bf S}_i+{\bf
  L}_i$ at each lattice site $i$ is conserved, recovering the familiar
conservation law of total angular momentum.  This means that the
amount of spin change must be matched exactly by the orbital angular
momentum change. What is less familiar, however, is the orbital angular
momentum change due to a laser field. With presence of a laser beam,
Eq. (\ref{orbital}) becomes \be \frac{d{\bf L}_i}{dt}=- \lambda {\bf
  L}_i\times {\bf S}_i-e {\bf E}(t)\times {\bf r}_i(t)
\label{orbital1}
\ee where ${\bf E}(t)$ is the laser electric field and ${\bf r}_i(t)$ is
the position which also depends on the laser field.  This extra term
is from the laser field and is the main course of the total angular
momentum {\bf J} change (see Fig. \ref{fig2}(a)).  However, the spin
does not have this extra term, so Eq. (\ref{spin}) remains valid even
with presence of laser. If we add Eqs. (\ref{spin}) and
(\ref{orbital1}) and move the orbital term to the right side, we find \be
\frac{d{\bf S}_i}{dt}=- \frac{d{\bf L}_i}{dt}-e {\bf E}(t)\times {\bf
  r}_i(t).
\label{spin2}
\ee Integrating Eq. (\ref{spin2}), we find \be {\bf S}_i(t)-{\bf
  S}_i(-\infty)=-\int_{-\infty}^t d{\bf L}_i-e \int_{-\infty}^t {\bf
  E}(t')\times {\bf r}_i(t') dt'. \label{com} \ee Since $-e{\bf E}(t)$
is the force on the electron and ${\bf r}_i(t)$ is the time-dependent
position, the last term in the above equation is the time-integrated
torque.  The amount of spin change is directly proportional to the
accumulated torque.  Therefore, a large spin change is possible for a
weak laser pulse as long as its pulse duration is long. This is the
essence of our theory.  The appendix has a more detailed discussion on
orbital angular momentum.

Next we show a numerical example to demonstrate many interesting
aspects of AOS. We choose a linearly polarized pulse along the $y$
axis, with the field amplitude of $0.09\rm V/\AA$. The field amplitude
is chosen to maximize the spin switching.  The photon energy is 1.6 eV
and the pulse duration is 60 fs. Initially, the spin points along the
$-z$ axis.  {The initial momentum of the electron is $9.11\times
10^{-26}$kgm/s, which corresponds to a velocity of 1$\rm \AA/fs$,
close to the Fermi velocity in metals, along the $z$ axis.}  Figure
  \ref{singlespin} shows three components of the total angular
  momentum {\bf J}, spin and orbital angular momentum as a function of
  time. Our laser pulse peaks at 0 fs. From Fig. \ref{singlespin}(a),
  one can see that before and after laser pulse excitation the total
  angular momentum ${\bf J}$ is conserved. Figure \ref{singlespin}(a)
  demonstrates clearly that the $z$ component $J_z$ starts out with
  $-1\hbar$ and ends around $+0.75 \hbar$, which occurs during
  interaction with the laser field. Different from $J_z$, $J_x$ is
  zero in the beginning, and as the laser field excites the system, it
  becomes negative. For our current spin and momentum configuration,
  $J_y$ is zero. To understand these features, we further plot the
  spin and orbital angular momenta separately. Figure
  \ref{singlespin}(b) shows that the orbital angular momentum $L_x$
  undergoes a rapid oscillation around zero after laser
  excitation. This is due to the laser-induced torque (the last term
  in Eq. \ref{orbital1}). Spin dynamics is different. We notice that
  $S_x$ undergoes a vertical shift to the $-x$ axis and then
  oscillates. After excitation, the spin and orbital momentum
  oscillate exactly out of phase because their spin-orbit torques
  differ by a negative sign.  This leads to the constant $J_x$ seen in
  Fig. \ref{singlespin}(a). Figure \ref{singlespin}(c) reveals the
  reason behind $J_y=0$, where $L_y$ and $S_y$ always oscillate out of
  phase with each other. We emphasize that this is the direct
  consequence of our current spin and momentum configuration. Whether
  the spin oscillates or not depends on both the laser parameters and
  the spin value (see another example below).

The $z$ components of the spin and orbital angular momentum are shown
in Fig. \ref{singlespin}(d). We see that $S_z$ switches from $-1\hbar$
to $0.75\hbar$, but $L_z$ only oscillates around zero. This is
consistent with the finding along the $x$ axis. The laser field
directly injects angular momentum into the orbital degree of freedom,
wherein both the $x$ and $z$ components undergo a rapid oscillation
around 0 fs because the position and momentum are directly influenced
by the laser field.  {Equation (\ref{orbital1}) shows that the
  orbital angular momentum change is subject to two separate
  contributions. Since the laser frequency is much higher than that of
  SOC, the rapid beating seen in the orbital angular momentum reflects
  that the laser field dominates the process over the spin-orbit
  coupling.  Equation (\ref{spin2}) shows that the rapid component in
  $d{\bf L}/dt$, which is a torque itself, compensates the torque due
  to the laser field. This compensation is the manifestation of the
  dipole selection rule that the photon angular momentum is absorbed
  by the electron's orbital angular momentum. In other words, the
  rapid oscillation is consumed by the orbital angular momentum as can
  be seen from Eq. (\ref{orbital1}).  The spin change is still from
  Eq. (\ref{spin}). Because the right-hand side is a cross-product, a
  $z$ component of the spin $S_z$ depends on the $x$ and $y$
  components of the spin and orbital angular momentum ($\tau_z=\lambda
  (L_xS_y-L_yS_x)$). Since we align the spin along the $z$ axis, both
  $S_x$ and $S_y$ are zero in the beginning. Therefore, to change
  $S_z$, one has to wait until $S_x$ and $S_y$ differ from zero, and
  the direct and rapid response seen in the orbital angular momentum
  does not occur to spins.  Spin has SU(2) symmetry and must interact
  with the laser field even times to be affected \cite{nc18}.  } This
is an important character of AOS. We will come back to this below.

\subsection{Effects of spin, momentum and laser-field 
  on spin switching}

In general, not all the spins in domain walls point in the same
direction in space.  Even illuminated by the same type of light, these
spins respond differently since their local quantization axes are
different.  A $z$-component for one spin could become an $x$ component
for another. Therefore, without considering spin configuration
explicitly, it is very difficult to develop a sound physical intuition
of spin reversal. Our study examines how spin switching depends on the
laser field direction.  Following the above discussion, we investigate
a case where both the initial spin ${\bf S}_0$ and initial momentum
${\bf P}_0$ point along the $-z$ axis. The direction of the laser
field ${\bf E}$ is very critical to AOS. For instance, if we align the
laser electric field and the initial spin along the same direction,
i.e., ${\bf E}\parallel {\bf S}_0\parallel {\bf P}_0$, then the
orbital angular momentum is going to be zero. From Eq.  (\ref{spin}),
we see the torque is zero, so the spin can not be reversed.

\begin{figure}
  \includegraphics[angle=270,width=1\columnwidth]{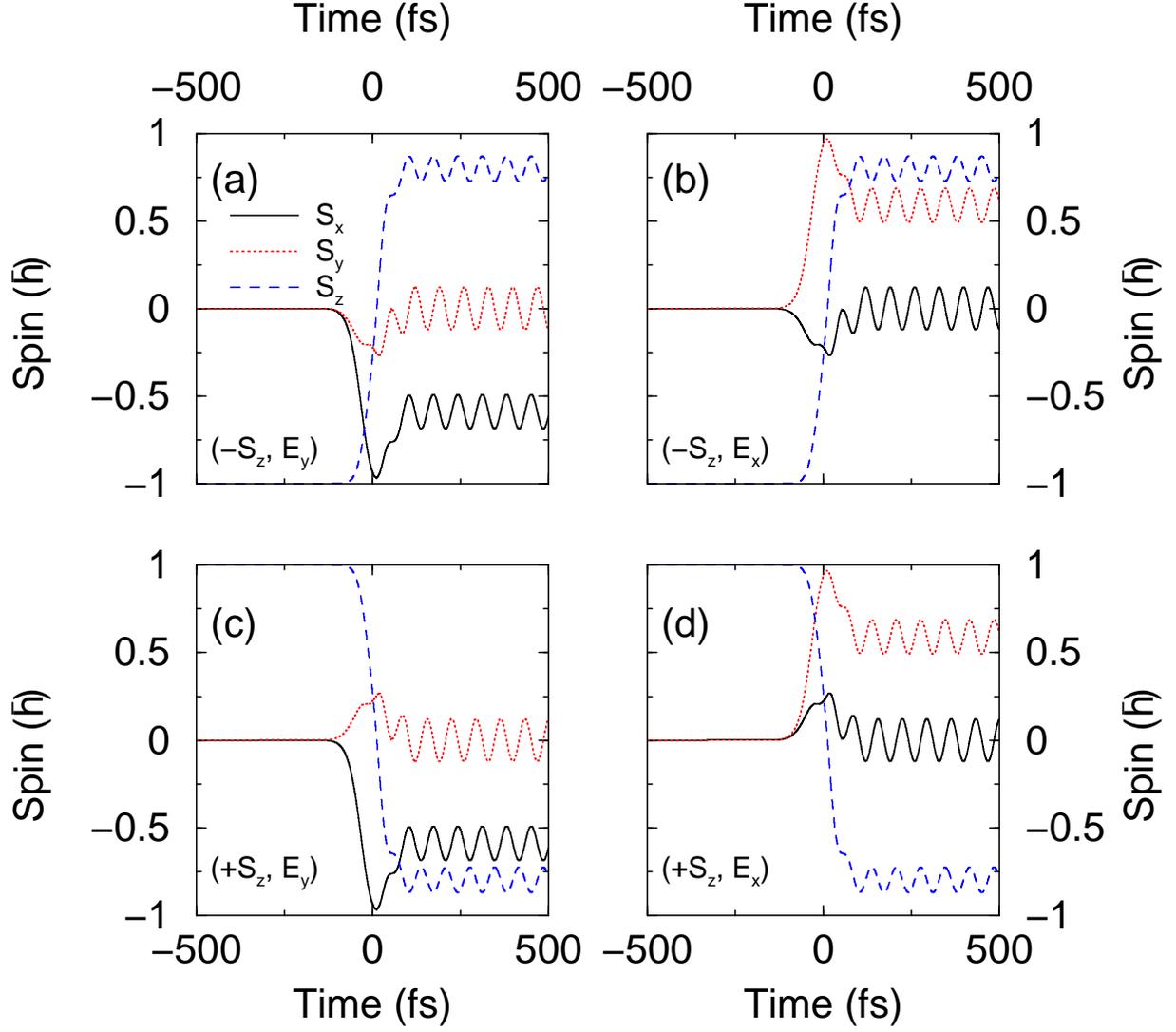}
  \caption{ Spin reversal under different electric field and spin
    configurations.  (a) Initial spin is along the $-z$ axis and the
    laser electric field ${\bf E}(t)$ is along the $y$ axis.  Spin
    components, $S_x$, $S_y$ and $S_z$, are denoted by the solid,
    dotted and dashed lines, respectively. The laser field amplitude
    is $0.09\rm V/\AA$ and the duration $T$ is 60 fs.  Spin $S_0=1\hbar$.
    (b) Initial spin is along the $-z$ axis but the electric field
    ${\bf E}(t)$ is along the $x$ axis.  (c) Initial spin is along the
    $+z$ axis and the electric field ${\bf E}(t)$ is along the $y$
    axis.  (d) Initial spin is along the $+z$ axis and the laser
    electric field ${\bf E}(t)$ is along the $x$ axis.  }
\label{fig3}
  \end{figure}

We next consider the laser field ${\bf E}$ along the $y$ axis,
perpendicular to ${\bf S}_0$ and ${\bf P}_0$.  Figure \ref{fig3}(a)
shows that three components of spin behave differently.  $S_x$
proceeds to a large negative value and oscillates, but $S_y$ shifts a
little and immediately oscillates after laser peaks. $S_z$ shows a
much stronger change and flips from the $-z$ axis to the $+z$ axis.
Now if we rotate ${\bf E}$ to the $x$ axis with respect to the $z$
axis by $-90^\circ$, the spin also rotates by $-90^\circ$. If we
compare Figs. \ref{fig3}(a) and \ref{fig3}(b), we can figure out the
transformation matrix as
   \begin{equation}
   Q_1=
  \left ( {\begin{array}{rrr}
  0 & -1  &~~0 \\
   1 &0  &0 \\
   0 & 0 &  1\\
  \end{array} } \right ).
\end{equation}

\begin{figure}
  \includegraphics[angle=270,width=1\columnwidth]{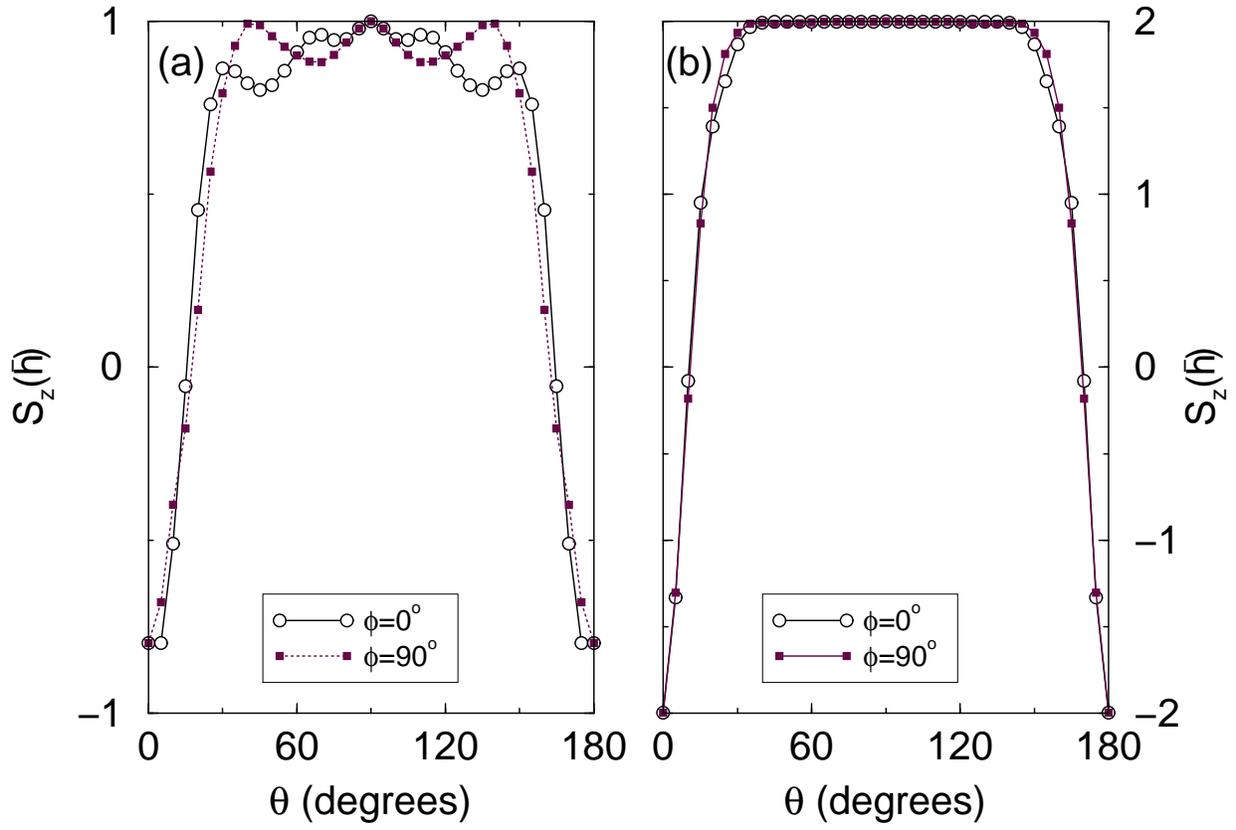}
  \caption{(a) Dependence of the final average spin on the polar angle
    $\theta$ of momentum ${\bf P}_0$ at two azimuthal angles
    $\phi=0^\circ$ and $90^\circ$ for $S_0=1\hbar$.  See
    Fig. \ref{fig1}(b) as how $\theta$ and $\phi$ are defined with
    respect to the spin.  The laser field amplitude is optimized at
    $A_0=0.09\rm V/\AA$, while the laser duration $T=60$ fs and photon
    energy $\hbar\omega=1.6$ eV. The spin reversal occurs only when
    ${\bf P}_0$ falls into a narrow angle around $+S_z$
    ($\theta=0^\circ$) or $-S_z$ ($\theta=180^\circ$).  (b) Dependence
    of final spin on the momentum ${\bf P}$ orientation for
    $S_0=2\hbar$.  The laser field amplitude is optimized at
    $A_0=0.20\rm V/\AA$. The rest of laser parameters are the same as
    (a). For a larger spin, the dependence is smoother because of a
    larger spin-orbit torque.  }
\label{fig4}
  \end{figure}

Next we keep {\bf E} along the $y$ axis but flip the spin from the
$-z$ to $+z$ axis. Figure \ref{fig3}(c) shows that {\bf S} undergoes a
180$^\circ$ rotation along the $x$ axis, with a different rotation
matrix
  \begin{equation}
   Q_2=
  \left ( {\begin{array}{rrr}
  1 &  0   &~~0 \\
   0 & -1&0 \\
   0 & 0 & -1\\
  \end{array} } \right ).
\end{equation}
One sees that the spatial orientation of spin determines how light
affects the spin. If we keep the spin along the $+z$ axis but rotate
{\bf E} to the $x$ axis as we did in Fig. \ref{fig3}(b), we find the
same transformation matrix $Q_1$ (Fig. \ref{fig3}(d)). To
comprehensively understand how the initial momentum ${\bf P}_0$
affects spin dynamics, we fix its magnitude, but change its spatial
orientation through $(\theta,\phi)$, whose definitions are presented
in Fig. \ref{fig1}(b).  We choose two different spin angular momenta
$S_0=1\hbar$ and $2\hbar$ since it is known the spin value affects how
laser switches spins \cite{epl16}.  Figure \ref{fig4}(a) plots the
final spins for $S_0=+1\hbar$ as a function of $\theta$ for
$\phi=0^\circ$ and $90^\circ$.  Figure \ref{fig4}(a) makes abundantly
clear that spatial orientation of ${\bf P}_0$ has a significant effect
on spin reversal.  The spin reverses only if $\theta$ is close to
$0^\circ$ or $180^\circ$ (see the large and negative final spins in
Fig. \ref{fig4}(a)).  $\phi$ has a small effect (compare the results
for $\phi=0^\circ$ and $180^\circ$), meaning that the transverse
momentum has no major effect. Consistent with prior results
\cite{epl16}, the spin itself also affects this dependence. In
Fig. \ref{fig4}(b), we use a larger spin of $2\hbar$, and we find that
the spin switchability is better, without a small oscillation seen in
Fig. \ref{fig4}(a), but the general conclusion remains the same.

\subsection{Switching uniform spins}

\begin{figure}

\includegraphics[angle=0,width=0.5\columnwidth]{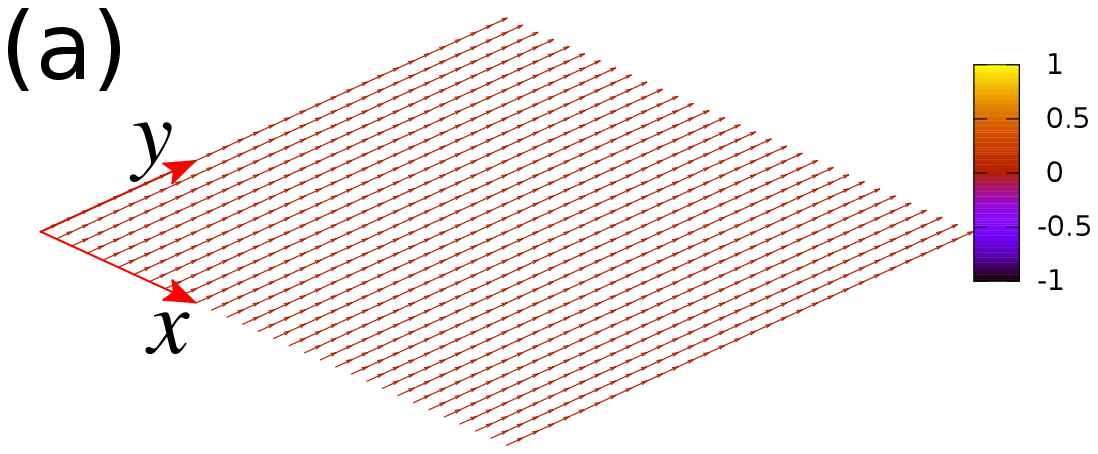}
\includegraphics[angle=0,width=0.5\columnwidth]{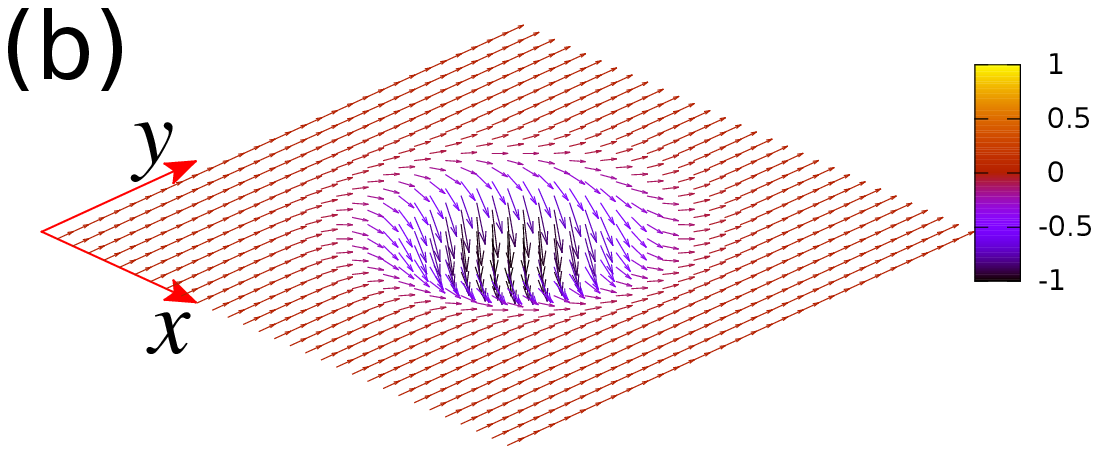}
\includegraphics[angle=0,width=0.5\columnwidth]{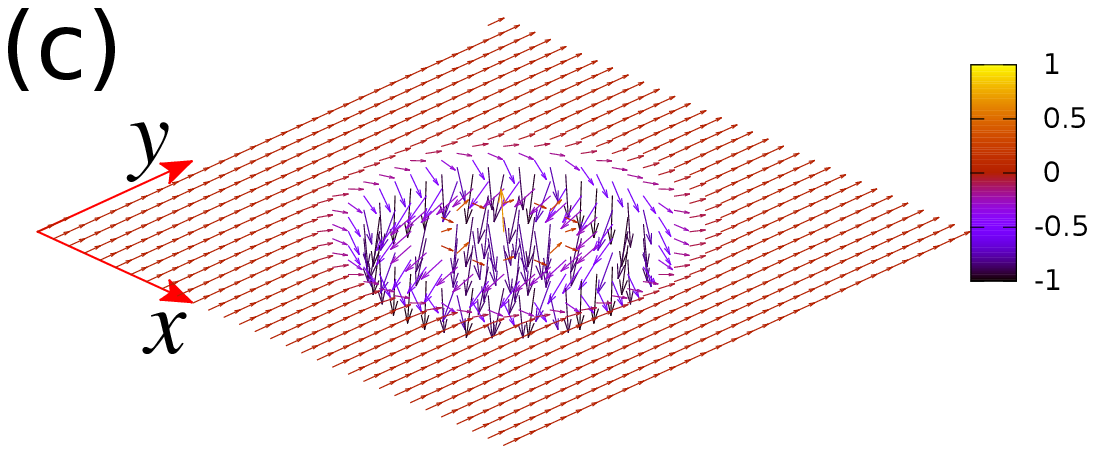}
\includegraphics[angle=0,width=0.5\columnwidth]{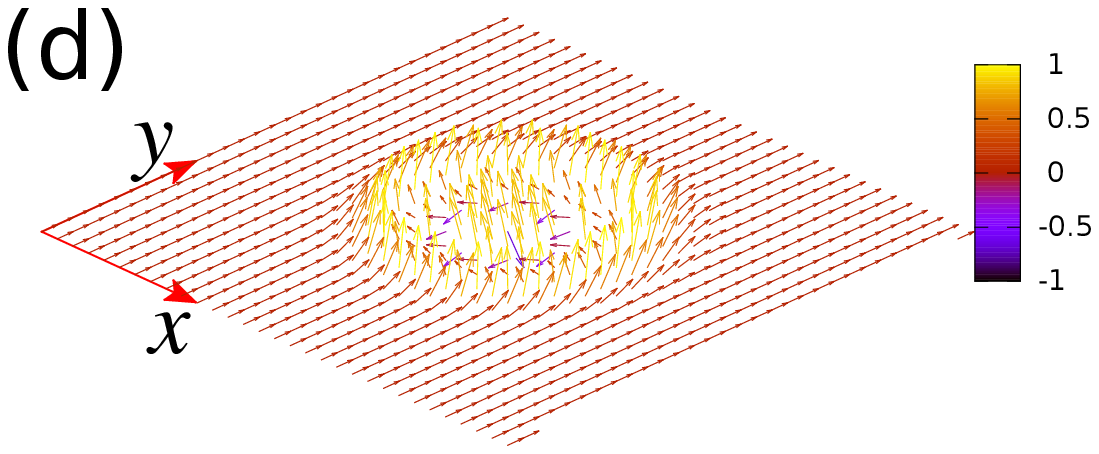}

  \caption{ Spin reversal in a uniform magnetized slab. (a) Initial
    spin configuration. All the spins are inplane and along the $y$
    axis. There are 501 spins along both the $x$ and $y$ axes. To
    reduce the huge data, we only show one spin every ten spins. (b),
    (c) and (d)
    Snapshot of spins at 123 fs under linearly polarized light,
    right- and left-circularly polarized light, respectively. Here
    Figs. (b)-(d) are revised. 
  }
\label{fig5}
  \end{figure}

In real magnetic materials, domains come with different shapes and
spin structures.  Different from the above single spin studies, spins
are exchange coupled together.  In the following, we investigate how
these different spin structures affect the spin reversal process.  We
start with  uniform spins initialized along the $y$ axis (see
Fig. \ref{fig5}(a)) with the same magnitude of $1\hbar$. The initial
momentum ${\bf P}_0$ is also in the $y$ direction, so we have an
optimal condition.  Following our finding in the single spin, we apply
a linearly polarized pulse along the $x$ axis, which allows a nonzero
orbital angular momentum. Spins precess in the time domain.  Figure
\ref{fig5}(b) shows a snapshot at 123 fs, after the laser pulse
peaks. This particular time is selected because it is after laser
excitation and its spin change is representative.  The color bar
denotes $S_z$ at each lattice site.  One sees that spin activities
concentrate around the center of the excitation where the laser field
is strongest. Spins clearly tilt toward the $-z$ axis and form a
dip. Around the circumference of the dip, spins smoothly transition
into the background. Note that to reduce the huge file size, we only
plot one out of every ten spins. When we switch to a right-circularly
polarized pulse ($\sigma^+$), we note a dramatic change. Figure
\ref{fig5}(c) reveals that the spin change becomes larger (compare
Figs. \ref{fig5}(b) and \ref{fig5}(c)). For instance, we notice that
spins in the active region point down and form a basin, and in the
center some spins are along the $x$ axis as well as the $-y$ axis,
i.e., inplane spin reversal, though the majority of spins point toward
the $-z$ axis. The situation is reversed if we use left-circularly
polarized light ($\sigma^-$). Figure \ref{fig5}(d) shows that spins
point along the $+z$ axis and form a mound. This is consistent with
our {theory} \cite{prb17} that $\sigma^+$ flips upspin down,
while $\sigma^-$ flips downspin up.

\subsection{Switching spins in a \neel wall}

\begin{figure}
\includegraphics[angle=0,width=0.5\columnwidth]{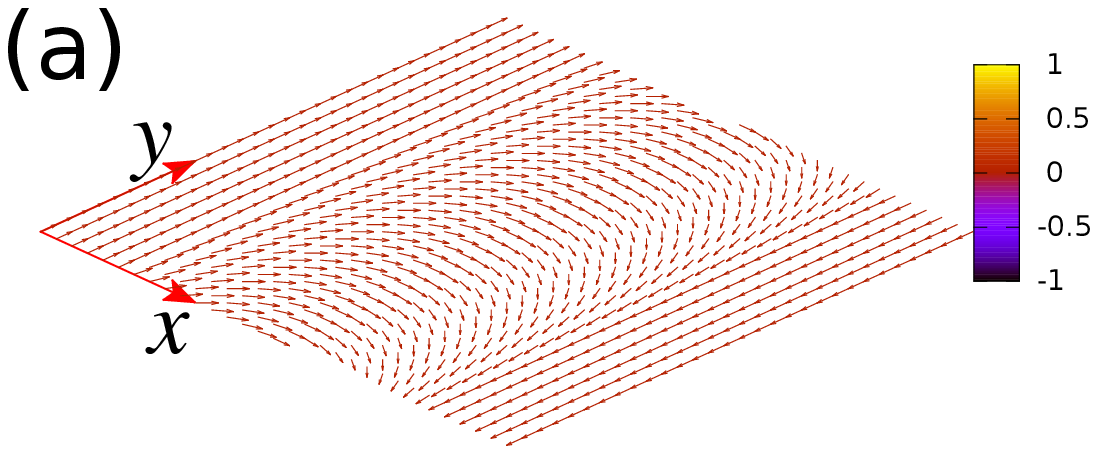}
\includegraphics[angle=0,width=0.5\columnwidth]{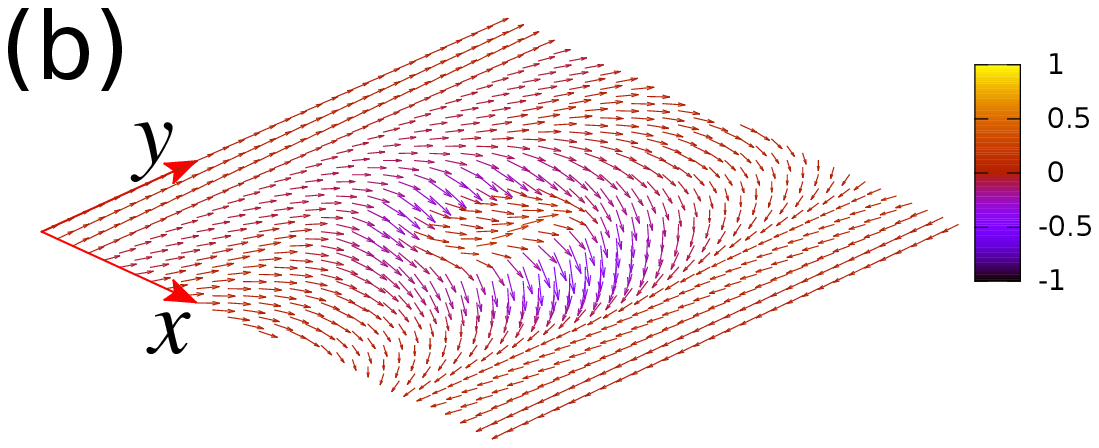}
\includegraphics[angle=0,width=0.5\columnwidth]{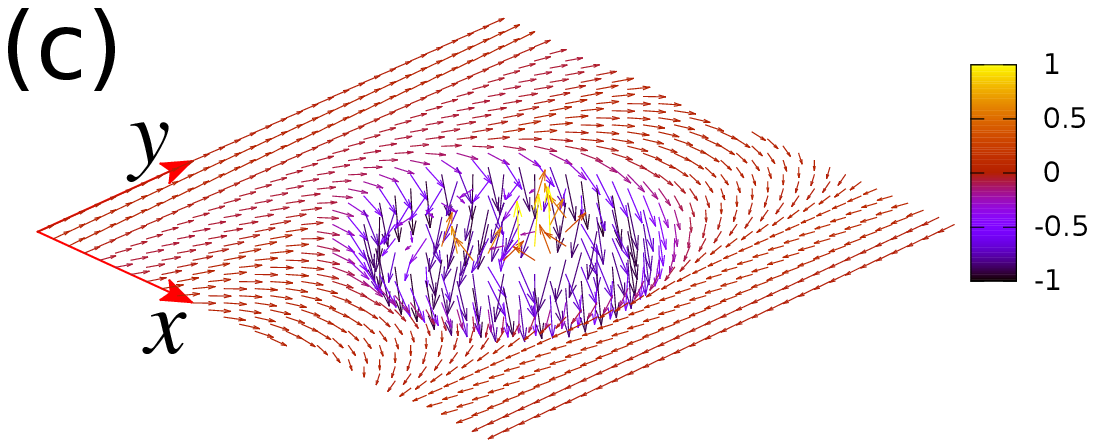}
\includegraphics[angle=0,width=0.5\columnwidth]{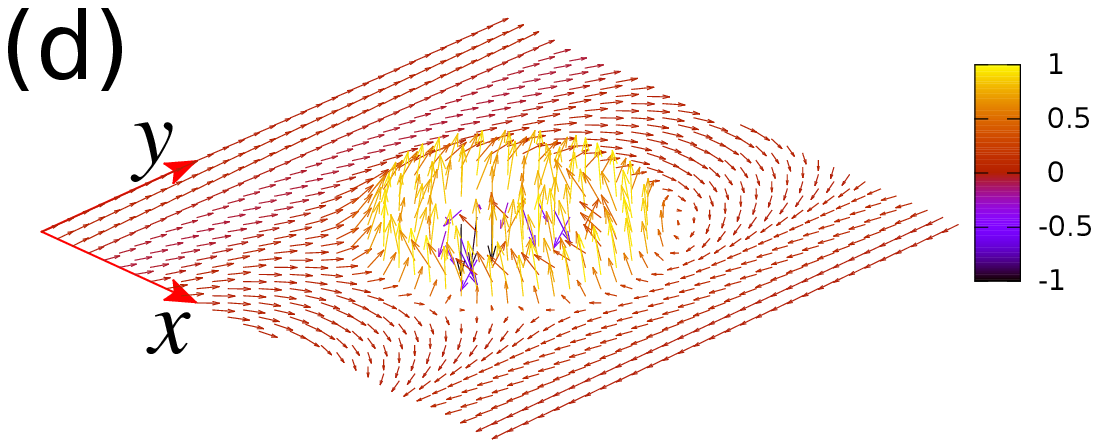}
\caption{Spin reversal across a \neel wall along the $y$ axis. (a)
  Initial spin configuration.  (b), (c) and (d) Snapshot of spins at
  123 fs under linearly polarized light, right- and left-circularly
  polarized light, respectively.  $\sigma^+$ tends to switch spin
  down, while $\sigma^-$ tends to switch spin up, which creates a
  basin (c) and mound (d) of spins, respectively. Around the
  vicinity of the basin and mound, spins are clearly reversed.  }
\label{fig6}
\end{figure}

A good place to understand the effect of the initial spin
configuration on AOS is to investigate what happens if the laser pulse
impinges on a \neel wall. Here, spins within the wall take different
spatial orientations, so locally each site has a different
quantization axis.  By studying their switchability, we can learn the
crucial connection between the spin and AOS. The wall is created by
two functions, $ S^x_i=S_0 \cos(\xi_i)$ and $S^y_i=S_0 \sin(\xi_i),$
where $S_0=1\hbar$, and $i$ is the lattice site index running from
$W_L$ to $W_R$.  $\xi_i=i\pi/(W_L-W_R)-\pi(W_L+W_R)/(W_L-W_R)/2$,
where $W_L$ and $W_R$ are the left and right limits of the wall,
respectively. $W_L$ and $W_R$ determine the width of the wall.  In our
case, the width is 200 sites.  This creates a \neel wall as illustrated
in Fig. \ref{fig6}(a).  Spins with lattice site indices less than
$W_L$ point in the $+y$ direction, and spins with lattice site indices
larger than $W_R$ point in the $-y$ direction. The second term in
$\xi_i$ ensures that spins in the middle of the wall point along the
$+x$ axis.

 We first subject the sample to a $\pi$ pulse. Figure \ref{fig6}(b)
 shows a snapshot at 123 fs. Interestingly, we find that the overall
 change is small. There is no major domain breakup. Only a small
 change is noticed in the center. The effect is even smaller than that
 in the uniform spin case. This indicates that the \neel wall can
 tolerate laser excitation more than the uniform spin slab.  {\clr The
   reason for this difference is that spin switching depends on the
   initial spin configuration, its relative orientation with respect
   to the laser electric field direction and the area of laser beam.
   The spins inside the \neel wall are parallel to the laser electric
   field, while the spins in the uniform spin slab are perpendicular
   to the laser electric field.  This difference leads to two
   different spin-orbit torques acted upon these two spin
   structures. For the spins outside the \neel wall, their directions
   are perpendicular to the laser electric field, so they are strongly
   affected if the laser beam radius is large, just as those in the
   uniform spin slab.}  The situation changes when we subject the
 magnetic slab to a $\sigma^+$ pulse (Fig. \ref{fig6}(c)), where the
 spins are switched to the $-z$ axis. What is even more interesting is
 that the spins around the perimeter of the pattern all point
 down. This is very similar to the experimental observation where the
 switched region appears around the circumference
 \cite{stanciu2007}. Figure \ref{fig6}(d) plots that $\sigma^-$
 creates a mixed spin domain, where the spins form a pattern tilting
 toward $+z$ axis and smoothly merging into the original \neel wall.

\begin{figure}
\includegraphics[angle=0,width=0.5\columnwidth]{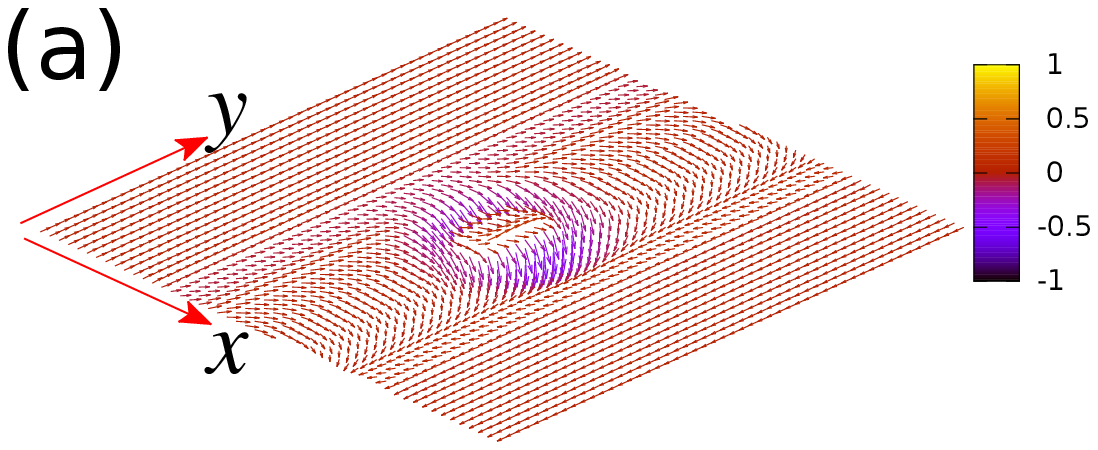}
\includegraphics[angle=0,width=0.5\columnwidth]{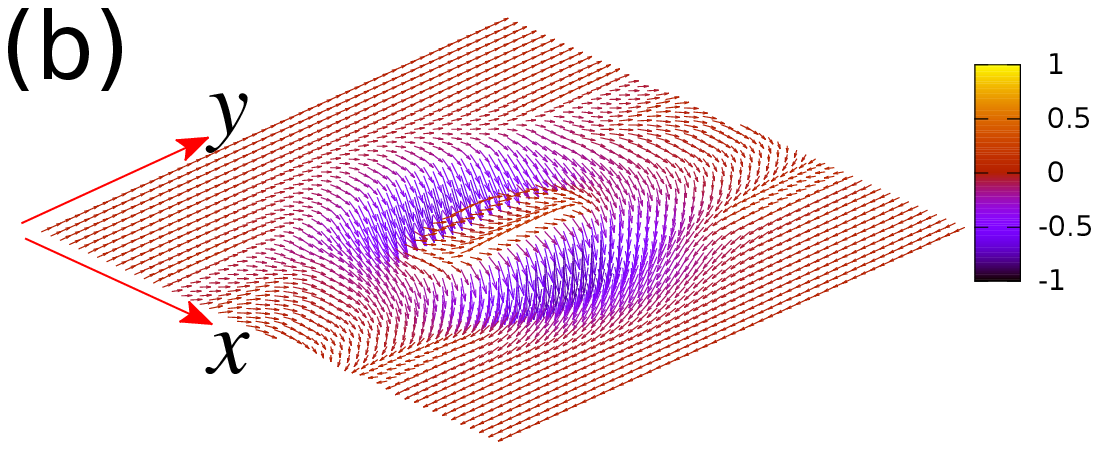}
\includegraphics[angle=0,width=0.5\columnwidth]{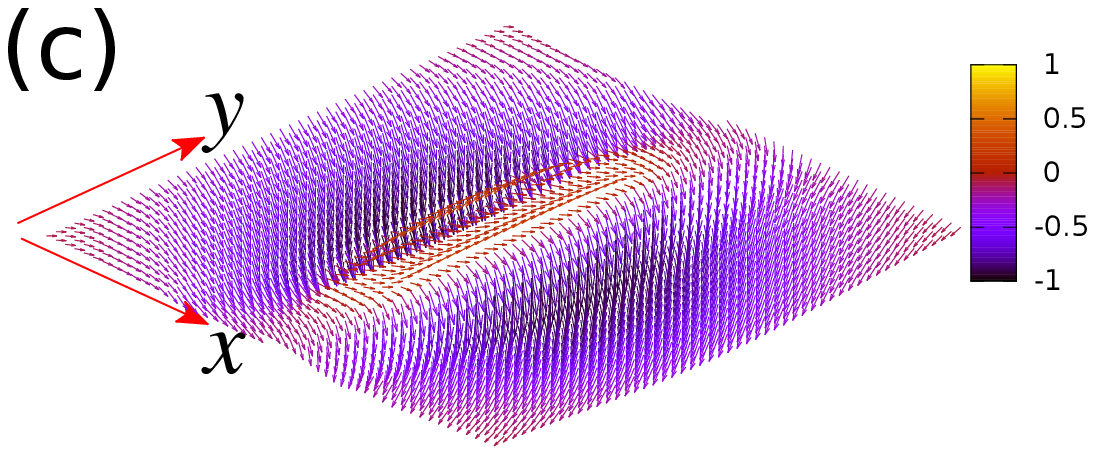}
\caption{Effect of laser beam radius on the spin switching in
  \neel walls under linearly polarized light along the $x$ axis.  (a)
  $R=50$. (b) $R=150$. (c) $R=200$.  It is clear that as $R$ increases,
  the affected region becomes larger. Interestingly, the original wall
  is still visible. Note that the image is on a much larger scale than
  that in Fig. \ref{fig6}.  }
\label{fig8}
\end{figure}

{We also investigate how the laser beam radius affects spin reversal
  in \neel walls. {\clr We apply a linearly polarized pulse along the
    $x$ axis}.  Figure \ref{fig8} shows results for three radii, (a)
  $R=50$, (b) $R=150$ and (c) $R=200$.  The image for $R=50$
  (Fig. \ref{fig8}(a)) is the same as Fig. \ref{fig6}(b), but on a
  larger scale with $501\times 501$, so the entire changed area
  appears to be smaller.  Figure \ref{fig8}(b) shows that as the
  radius becomes larger, the affected region is larger. However, the
  spin in the \neel wall is protected, with a very small change. The
  main change occurs in the vicinity of the excitation center, where
  the spins are along the $y$ or $-y$ axis and experience a stronger
  torque. This spatial inhomogeneity persists even if we use $R=200$
  which matches the width of the wall (see Fig. \ref{fig8}(c)).} {\clr
  Although spins outside the \neel wall switch roughly to the same
  direction, spins inside the \neel wall do not.  This is because
  spin-orbit torques on those spins inside the \neel wall are
  different from those outside the \neel wall.  Spins inside the \neel
  wall do not orient in the same direction, and they mainly lie along
  the $x$ axis, so the domain wall retains.  }

\subsection{Switching spins in a Bloch wall}

\begin{figure}
\includegraphics[angle=0,width=0.5\columnwidth]{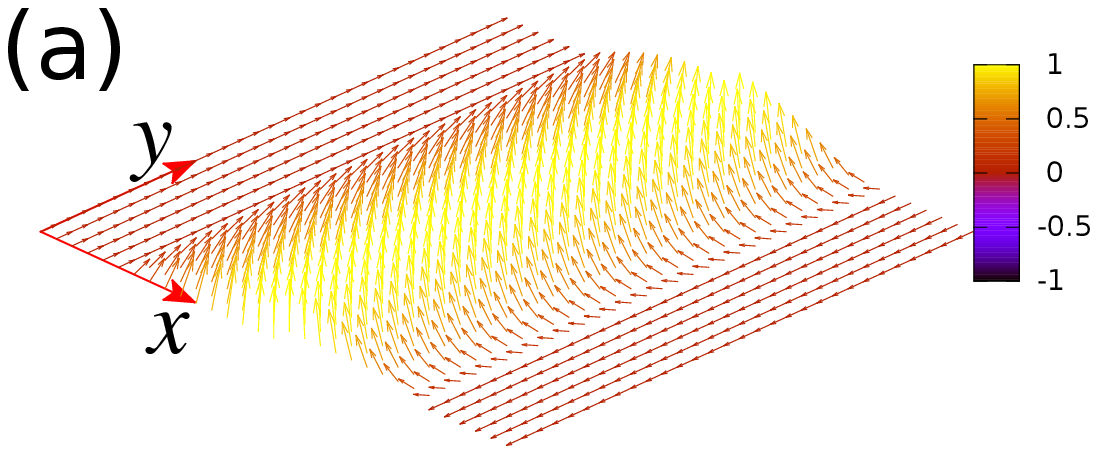}
\includegraphics[angle=0,width=0.5\columnwidth]{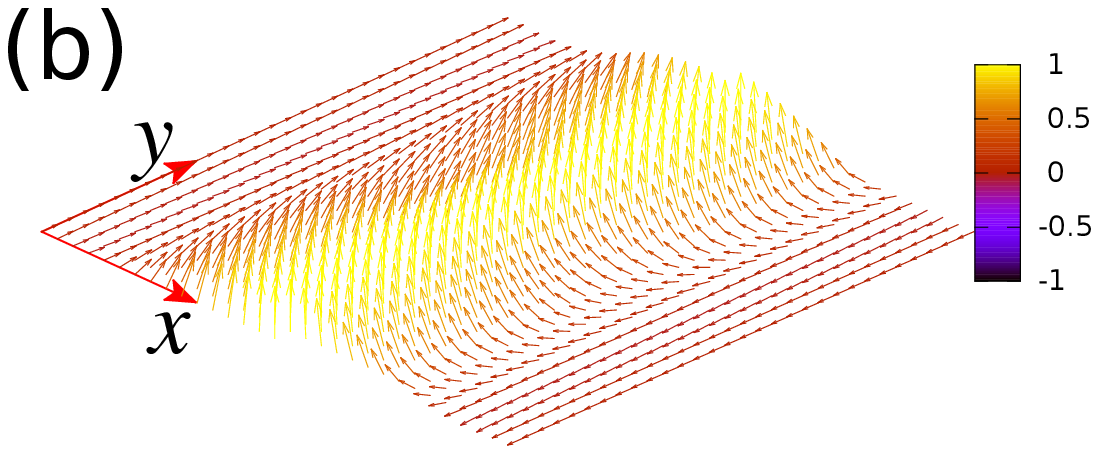}
\includegraphics[angle=0,width=0.5\columnwidth]{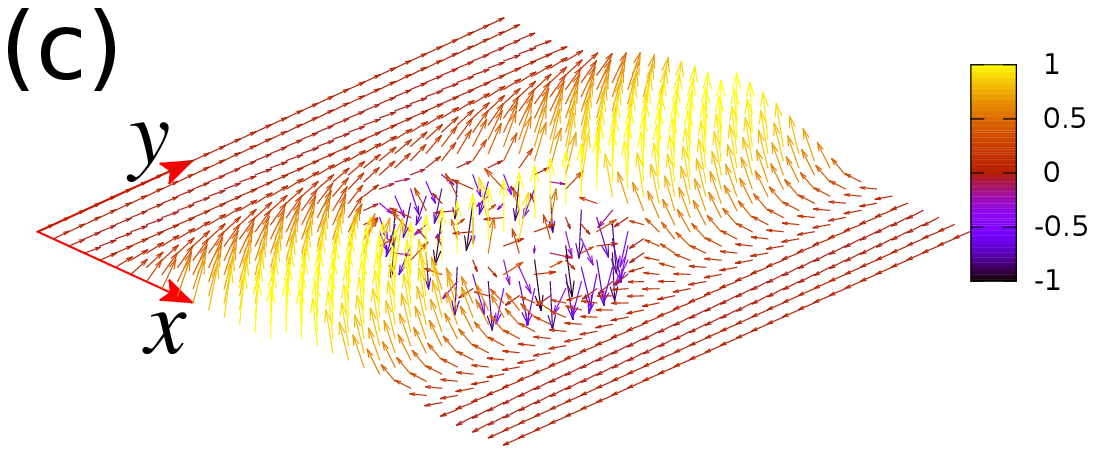}
\includegraphics[angle=0,width=0.5\columnwidth]{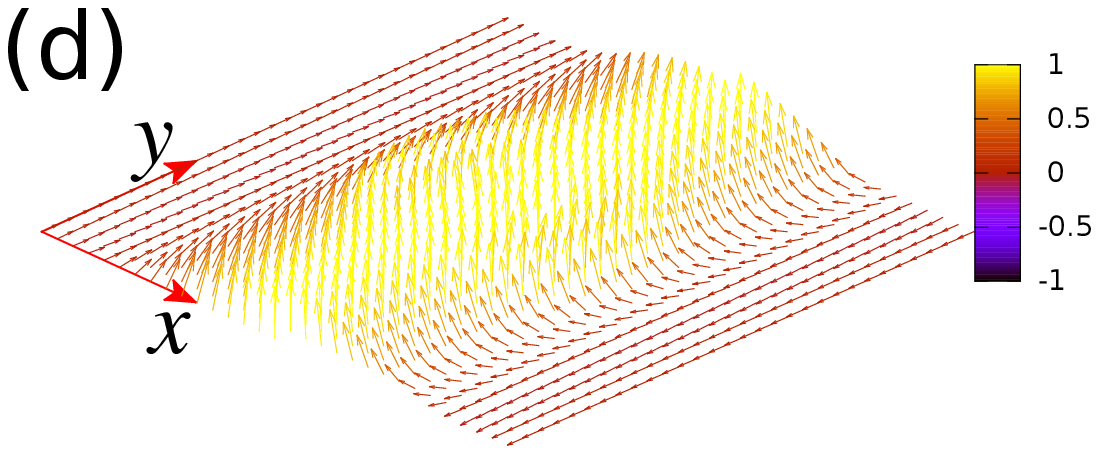}
\caption{Spin reversal across a Bloch wall along the $y$ axis. (a)
  Initial spin configuration.  (b), (c) and (d) Snapshot of spins at
  123 fs under linearly polarized light, right- and left-circularly
  polarized light, respectively. A dramatic impact on the wall is from
  the $\sigma^+$ light, where the domain wall is disrupted. Changes
  due to the $\sigma^-$ light are small due to its helicity
  \cite{prb17}.  }
\label{fig7}
\end{figure}

To this end, the initial spins are in the same plane of the light
polarization. In the Bloch wall, spins are along the normal
direction. Our domain wall width also has 200 lattice sites.  The wall
is created by two functions, $ S^y_i=S_0 \sin(\xi_i)$ and $S^z_i=S_0
\cos(\xi_i),$ which smoothly merges into the neighboring domains. The
produced spin structure can be seen from Fig. \ref{fig7}(a). We start
with a $\pi$ pulse. The results are shown in Fig. \ref{fig7}(b). We
see that the effect on the domain wall is very weak. This is similar
to the \neel wall (Fig. \ref{fig6}(b)). By contrast, a $\sigma^+$ pulse
has a much stronger effect. Recall that a $\sigma^+$ pulse switches
spin from up to down \cite{prb17}, so the current spin configuration
is an ideal place to play a strong role. Here the spin dynamics is
violent, with a distinctive feature of spin reversal around the
vicinity of the excitation region.  In the center spins are much more
chaotic. Beyond the excitation center, the domain structure remains
intact.  We note that the radius of the laser spot is 50 lattice
sites.  This well localized spin structure of about 200 sites
represents a possibility to create a much small magnetic domain,
potentially very useful for development of denser magnetic storage
devices. We also investigate how $\sigma^-$ affects the Bloch
wall. Figure \ref{fig7}(d) shows a snapshot at 123 fs that the spins
are not strongly affected. This is because $\sigma^-$ tends to flip
spins from down to up, but once spins are already up, the effect on
them is very small \cite{prb17}.

We can make a connection to existing experiments. Stanciu \et
\cite{stanciu2007} showed that spin switching occurs at the vicinity
of the demagnetized area. From our study, we indeed observe that the
reversed spins mainly appear around the ring of the excitation (see
Figs. \ref{fig6}(c) and \ref{fig6}(d) as well as Figs. \ref{fig7}(c)
and \ref{fig7}(d)). Note that Stanciu's sample is ferrimagnetic and in
principle two different spin sublattices play a role.  Our findings
serve a theoretical basis for these experimental results and are
expected to have important impacts in the future.

\begin{figure}
\includegraphics[angle=0,width=1\columnwidth]{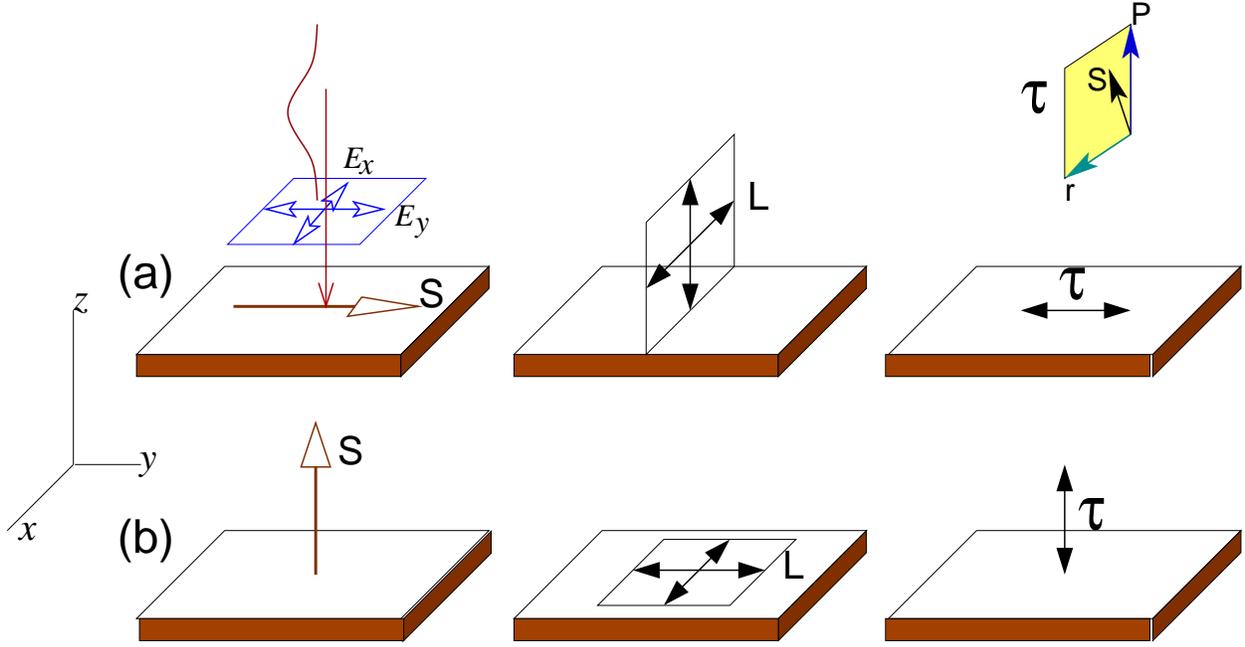}
\caption{Basic physics of all-optical spin switching. Here a
  laser pulse impinges a sample at normal incidence along the $-z$
  axis, so the electric field falls in the $xy$ plane.  (a) Spin is
  in plane along the $y$ axis. Middle: Plane of orbital angular
  momentum ${\bf L}$; Right: Spin-orbit torque is along the $y$
  axis. (Top right): Relation among {\bf S}, {\bf r}, {\bf P}, and
  ${\tau}$.  (b) Spin is out of plane along the $z$ axis. Middle:
  Plane of orbital angular momentum ${\bf L}$ is in the $xy$ plane;
  Right: Spin-orbit torque is along the $z$ axis.  }
\label{fig9}
\end{figure}

{

\subsection{Spin-orbit torque theory}


The above numerical results present us rich information how spins are
affected with different spin configurations. We can reveal further
insight into the underlying physics by investigating how the orbital
angular momentum enters the picture.  We seek the guidance from the
spin-orbit torque $\tau=\lambda ({\bf L} \times {\bf S})$.  To be
definitive, consider a spin inplane and along the $y$ axis as
illustrated in Fig. \ref{fig9}(a), with the laser pulse propagating
along the $-z$ axis, normal incidence. The laser electric field is
decomposed along the $x$ and $y$ axes, respectively.  To reverse spin
$S_y$, one must have a nonzero torque or $\tau_y\ne 0$. Because
$\tau_y=\lambda (L_z S_x-L_xS_z)$, $L_z$ and $L_x$ can not both be
zero.  As illustrated in Fig. \ref{fig4}, to maximize AOS, in our
simulation we set the initial momentum along the $y$ axis, same as the
spin. For this inplane spin configuration, only the $x$ component
$E_x$ of the laser field is effective.  If the laser polarization is
also along the $y$ axis (only $E_y$ is nonzero), then the electron
only moves along the $y$ axis, generating a zero orbital angular
momentum due to ${\bf L}={\bf r}\times {\bf P}$. Once the orbital
angular momentum is zero, the torque is zero, so there is no spin
switching.  The middle figure of Fig. \ref{fig9}(a) shows that the
plane of the orbital angular momentum is perpendicular to the spin, so
the torque $\tau$ in the right figure is in the same direction of the
spin, so switching is possible.

Figure \ref{fig9}(b) plots the spin pointing out of plane
\cite{jpcm17b}. In this case, the orbital angular momentum plane is in
the $xy$ plane (see the middle figure in Fig. \ref{fig9}(b)).  Because
$\tau_z=\lambda (L_x S_y-L_yS_x)$, $L_x$ and $L_y$ can not both be
zero if switching is desired.  The inplane orbital angular momentum
ensures the resultant torque is along the spin direction (see the
right figure in Fig. \ref{fig9}(b)) to switch spins. These two special
cases for the spin orientation point out a crucial fact that to switch
the spin the orbital angular momentum plane must be perpendicular to
the spin.

To see how geometrically the electron position, momentum and spin are
intertwined, we rewrite the torque as a triple vector product
$\tau=\lambda ({\bf r}\times {\bf P}) \times {\bf S}$, \be
\tau=\lambda({\bf r}\cdot {\bf S}) {\bf P}-({\bf P} \cdot {\bf S})
            {\bf r}, \ee which highlights that the torque lies on the
            plane defined by the momentum and position (see the upper
            right figure in Fig. \ref{fig9}).  If a spin happens to
            lie in this plane, it will be strongly affected.  This
            explains why in Fig. \ref{fig4} to switch spin
            effectively, the angle between the spin and momentum and
            that between the spin and position must remain small.
            More importantly we find the module squared of $|\tau|^2$
            can be written as \be |\tau|^2=\lambda^2 |{\bf S}|^2|{\bf
              r}|^2|{\bf P}|^2(\cos^2\alpha + \cos^2\beta -2\cos\alpha
            \cos\beta\cos\gamma)
\label{replyeq1} \ee where $\alpha$ is the angle between {\bf r} and
    {\bf S}, $\beta$ is between {\bf P} and {\bf S}, and $\gamma$ is
    between {\bf r} and {\bf P}.  Here we see that the dependence of
    the spin-orbit torque on the position and momentum is similar as
    expected. The smaller the angle is, the larger the torque is.  We
    note that these three angles are not independent. For instance, if
    $\gamma=0^\circ$, then $\alpha=\beta$.  There is an internal
    competition between the position and momentum: To have a large
    torque, the spin has to lie as close as possible to both the
    position and momentum, but the position and momentum must not
    align. This competition underlies the theory of spin-orbit torque
    for all-optical spin switching.
}

 \section{Conclusion}

We have shown that laser-induced all-optical spin switching
sensitively depends on both the intrinsic properties such as spin
structure and initial momentum of electron and the extrinsic laser
parameters such as laser polarization. We find that while for a single
spin, the total angular momentum is always conserved in the absence of
a laser field, a laser pulse fundamentally alters the relation between
spin and orbital angular momenta. At least within our current model,
the angular momentum transfer between the spin and orbital does not
constitute a major channel for spin switching. Instead, the
time-integrated spin-orbit torque is the main driver for spin
change. Because of this integration, spins can be excited even with a
weak laser pulse, provided that the pulse is long enough. We further
show that the laser field polarization directly influences the spin
vector. Rotating the polarization from the $y$ axis to $x$ axis also
rotates the spin vector. However, if we flip the initial spin, the
same laser field induces a different spin precession. These findings
are manifested in a much larger system. For a magnetic thin film with
inplane magnetization, linear ($\pi$), right and left circularly
polarized light ($\sigma^+$ and $\sigma^-$) induce different
changes. $\sigma$ light generates a distinctive pattern of spin
reversal, while $\pi$ light only tilts spins. When a light field of
the same kind illuminates \neel and Bloch walls, the resultant spin
patterns have imprints of the starting domains. Both the \neel and
Bloch walls can withstand $\pi$ light, but suffer a dramatic spin
change from $\sigma$ light.  $\sigma^+$ and $\sigma^-$, both being
circular, also have a different effect on the spin. In Bloch walls
with the spins pointing along the $+z$ axis, $\sigma^-$ has little
effect on the wall, but $\sigma^+$ reverses a major portion of the
domain. These findings present a case that can be experimentally
tested. Investigation of different spin structures has been a focus
recently. Several prior studies have even proposed the domain size
\cite{elhadri2016} as a criterion for the all-optical spin
switching. We believe that by examining different spin structures, one
can learn more about the interplay between spin, orbital and laser
field, which is potentially important for future development of fast
magnetic storage.

 \acknowledgments
  This work was solely supported by the
U.S. Department of Energy under Contract No. DE-FG02-06ER46304. Part
of the work was done on Indiana State University's high performance
quantum and obsidian clusters.  The research used resources of the
National Energy Research Scientific Computing Center, which is
supported by the Office of Science of the U.S. Department of Energy
under Contract No. DE-AC02-05CH11231.

$^{*}$ guo-ping.zhang@outlook.com

\appendix

\section{Orbital angular momentum}

At the center of all-optical spin switching is the orbital angular
momentum.  First-principles calculations in solids often work in the
crystal momentum space. Due to the translational symmetry in solids,
the orbital angular momentum is largely quenched. When a laser pulse
excites the system, such a calculation surely underestimates the
orbital momentum change. To see this clearly, consider a single atom
subject to a laser pulse. The atom must exchange its orbital angular
momentum $L$ with the light field by $\pm 1 \hbar$, i.e. the selection
rule $\Delta l =\pm 1$, where $l$ is the orbital angular momentum
quantum number. However, for solids, the common wisdom is that the
orbital angular momentum $\langle L_z \rangle$ is largely quenched
\cite{prb08} because for every positive $m_l$ there is a negative
$m_l$. Then the summation over $l$ is \be \sum_l \langle L_z \rangle
=0~. \label{eq1} \ee Thus, a solid does not allow a big change in
$L$.

To reconcile the difference between the optical selection rule and the
degeneracy due to translational symmetry, it is often assumed that the
increase in orbital momentum has to be quenched immediately
\cite{boeglin2010,tows2015}. Fortunately, this paradoxical
contradiction does not arise for three reasons. First, only a small
portion of a sample is exposed to laser excitation, so the
translational symmetry is broken and the crystal momentum is no longer
a good quantum number, which favors a real space view.  Second, under
laser excitation, original degenerate states, which lead to the
orbital momentum quenching, are no longer degenerate. The summation
over degenerate states in Eq. (\ref{eq1}) is no longer necessary.
Third, optical excitation does not only affect the longitudinal
component of the orbital angular momentum $L_z$, but also the
transverse ones $L_x$ and $L_y$. To reverse spins, transverse
components are more important because they constitute the spin-orbit
coupling that reverses spins.  Neither $L_x$ nor $L_y$ is probed by
x-ray magnetic circular dichroism (XMCD), so prior time-resolved XMCD
results can not be used as evidence for the insignificance of the
orbital angular momentum.  More importantly which component is
longitudinal or transverse depends on the quantization axis.  The
reason why our model is successful is mainly because the initial
orbital angular momentum is treated properly. We work in the real
space, so the geometry of laser excitation is taken into account
realistically.

\end{document}